\documentclass[11pt,5p,twocolumn,preprint]{elsarticle}

\journal{~}

\usepackage[utf8]{inputenc}		%
\usepackage[T1]{fontenc}		%
\usepackage{graphicx}			%
\usepackage{subcaption}
\usepackage{amssymb} 			
\usepackage{amsmath} 			
\usepackage{wasysym} 
\usepackage{array}
\usepackage{tabularx}
\usepackage{multirow}
\usepackage{lineno}

\begin{document}

\begin{frontmatter}

\title{Towards a Physically Motivated Planetary Accounting Framework}

\author{M. Barbosa}
\ead{up201305930@fc.up.pt }

\author{O. Bertolami}
\ead{orfeu.bertolami@fc.up.pt}

\author{F. Francisco}
\ead{frederico.francisco@fc.up.pt}

\address{Departamento de Física e Astronomia and Centro de Física do Porto, Faculdade de Ciências, Universidade do Porto, Rua do Campo Alegre 687, 4169-007 Porto, Portugal}

\begin{abstract}
 In this work we present a physically motivated planetary Accounting Framework for the Earth System. We show that the impact of the human activity in terms of the Planetary Boundary variables can be accounted for in our Landau-Ginzburg phase transition physical formulation. We then use the interaction between climate change and ocean acidification mechanisms to exemplify the relation of the concentration and flux of substances of the Planetary Boundaries variables, as proposed by the accounting framework of Kate and Newman, with the underlying thermodynamical transformation, quantifiable by the Landau-Ginzburg inspired model. In this work we present a physically motivated planetary Accounting Framework for the Earth System. We show that the impact of the human activity in terms of the Planetary Boundary variables can be accounted for in our Landau-Ginzburg phase transition physical formulation. We then use the interaction between climate change and ocean acidification mechanisms to exemplify the relation of the concentration and flux of substances of the Planetary Boundaries variables, as proposed by the accounting framework of Kate and Newman, with the underlying thermodynamical transformation, quantifiable by the Landau-Ginzburg inspired model.
\end{abstract}


\end{frontmatter}


\section{Introduction}

The impact of human activities on the Earth System (ES) has become a defining issue of our time and has given origin to the proposal of a new geological epoch, the Anthropocene. Given the extent of human action, measuring its impact in an objective way is crucial to ensure that fair and rational stewardship measures are implemented in the near future. The mounting evidence, not only of anthropogenic climate change, but also of the deterioration of several ecosystems makes it urgent. Ideally, consensual global actions along the lines of the 1987 Montreal Protocol to halt the destruction of the ozone layer and the 1997 Kyoto Protocol for reducing the greenhouse-gas emissions, renewed in Paris in 2015, would be agreed, allowing for the design of strategies for a sustainable future with quality of life all humankind. A fair, reliable and uncontroversial accounting system for human impacts is an indispensable tool for any such effort.

Accounting for the impact of the human action on the ES has been the purpose of several proposals such as the $I=PAT$ measure \cite{Ehrlich:1971}, the Kaya identity \cite{Kaya:1993} and the well developed Ecological Footprint proposed in the 1990's \cite{Rees:1994} which since 2003 is being carried out in a systematic way by the Global Footprint Network. Although these proposals have several virtues, as centred on the socio-economical nature of human activities, they lack an unequivocal correspondence to the natural biogeochemical and physical processes of the ES that are now being disrupted. For instance, the quite accomplished Ecological Footprint is based on the confrontation of a measure of bio-capacity for a given territory, measured in global hectares per capita, against the overall consumption of energy, biomass, building material, water, etc, measured in the same units. Of course, this involves a quite complex conversion methodology, whose undeniable usefulness contrasts with its somewhat arbitrary nature, which in turn makes the connection with the impact on the ES somewhat indirect and opaque.
 
In order to achieve a more direct method to access the impact of the human action on the ES, one should start from the very basis of its functioning principles, as proposed in Ref. \cite{Steffen:2011}.

Recently, we have proposed that the ES transition from the Holocene to the Anthropocene could be regarded as a phase transition and described by the Landau-Ginzburg theory \cite{Bertolami:2018a}. The relevant thermodynamic variable of this approach is the free energy, $F$, and it was suggested that the relevant order parameter, $\psi$, of the description is the relative temperature deviation from the Holocene average temperature $T_{\rm H}$, that is $\psi = (T -  T_{\rm H})/T_{\rm H}$.

This framework allows for establishing the state of the ES in terms of the relevant physical variables, $(\eta, H)$, where $\eta$ corresponds to the astronomical, geophysical and internal dynamical effects while $H$ stands for the human activities.

The proposed physical model allows for obtaining the so-called Anthropocene equation, \textit{i.e.}, the evolution equation of the ES once it is dominated by the human activities, and to show that the transition from the Holocene conditions to the Anthropocene arises from the great acceleration of the human activities that was witnessed by the second half of the 20th century \cite{Steffen:2014}.

It has also been shown how the human activities function, $H$, can be decomposed into its multiple components, with a straightforward correspondence to the measurement of human impacts in terms of the Planetary Boundaries \cite{Steffen:2015}.

This logic has recently been used to set up an accounting framework that manages to gauge the human impact in terms of 10 quotas based on the 9 planetary boundary parameters \cite{Meyer:2018}. This procedure allows for an empirical basis for the environmental issues and a set of accessible indicators that can be adopted at various scales by various agents, being thus a poly-scalar approach.

Subsequently, we performed a phase space analysis of temperature field, $(\psi, \dot{\psi})$ and showed that the recently discussed Hothouse Earth scenario, \cite{Steffen:2018}, corresponds to a stable minimum and, therefore, to an attractor of the trajectories of the dynamical system that describes the ES \cite{Bertolami:2018b}. 

The purpose of the present work is to provide a physical support to the accounting framework of Ref. \cite{Meyer:2018}. This paper has the following structure: in the next section we review our Landau-Ginzburg model proposal and discuss its main implications for the description of the ES. In section 3, we discuss how after splitting the human activity on its impact on the parameters of the planetary boundaries and disregarding interaction terms, our physical framework can naturally give origin to an accounting framework that resembles up to multiplicative constants the framework proposed in Ref.\,\cite{Meyer:2018}. In fact, it is possible to show, given the generality of our approach, that it contains, for instance, under certain conditions, the $I=PAT$ proposal. However, a distinct feature of our approach is that it allows for considering the interaction terms between the various planetary boundaries, implying that any accounting framework is useful provided these interacting terms are neither important nor evolving in a time scale such that the effect of these terms can be relevant. In section 4 we consider the interaction between two planetary boundaries, namely the concentration of atmospheric carbon dioxide and the ocean acidity. In section 5 we present our results. Finally, in section 6 we discusss our conclusions.


\section{The Physical Model}

The fundamental insight of our physical model of the ES is its description of the transition from the Holocene to the Anthropocene as a phase-transition through the Landau-Ginzburg Theory \cite{Bertolami:2018a}. The main thermodynamic variable is the free energy,
\begin{equation}
	F(\eta,H) = F_0 + a(\eta)\psi^2 + b(\eta)\psi^4 - h(\eta)H\psi,
	\label{eqn:FreeEnergy}
\end{equation}
where coefficients $a(\eta)$, $b(\eta)$ and $h(\eta)$ depend on the set of variables $\eta$ and the effect of human intervention, $H$, is introduced as an external field modelled as linear term in the order parameter. The effect of this term is a clearly destabilizing one, as it is easy to show that the model predicts that, in the Anthropocene,
\begin{equation}
	\langle \psi \rangle \approx \left(\frac{H}{4b}\right)^{1/3},
\end{equation}
which means the equilibrium temperature of the ES, represented by $\langle \psi \rangle$, grows proportionally to the cubic root of human activities \cite{Bertolami:2018a}.

It has also been shown that the dynamical model arising from this description has an attractor of trajectories, provided that human action, $H$, is finite and some damping is introduced into the ES model. This critical point of the dynamical system corresponds necessarily to an ES trajectory towards a minimum where the temperature is greater than during the Holocene equilibrium \cite{Bertolami:2018b}.  

It is relevant to point out that these results are consistent with the features of the ES trajectories in the Anthropocene arising from the qualitative discussion of Ref.\,\cite{Steffen:2018}. Clearly, this means an increase in the global temperature which can lead to a chain failure of the main regulatory ecosystems of the ES, that already show tipping point features \cite{Steffen:2018}. 

Given that the purpose of an accounting framework is to gauge the impact of the human drivers, $H$, with respect to the Holocene conditions expressed by the first three terms in the free energy, we can isolate the term describing human activities as the change in the ES free energy in the Anthropocene,
\begin{equation}
	\Delta F(H)|_\text{Anthro} = - H\psi,
\end{equation}
where we have absorbed the constant, $h$, into the definition of $H$. This can be safely done because the time scale of human impacts is so much faster than that of natural driven change to the ES.

The effect of human activities in altering the optimum Holocene conditions can be decomposed into its different components, denoted by $h_i$. These components have a straightforward correspondence to the set of 9 or 10 parameters that compose the Planetary Boundaries (PB) framework \cite{Steffen:2015}. In the context of the PB, these parameters specify the state of the ES in terms of its deviation from the Holocene-like conditions that sustain human civilisation as we know it. The maintenance of these conditions is ensured if the ES remains within the so-called Safe Operating Space (SOS), which sets a limit on each of the PB components \cite{Rockstrom:2009}. The function $H$ describing human intervention can thus be written as \cite{Bertolami:2018b}:
\begin{equation}
	H = \sum_{i=1}^{9} h_i + \sum_{i,j=1}^{9} g_{ij} h_i h_j + \sum_{i,j,k=1}^{9} \alpha_{ijk} h_i h_j h_k + \ldots,
	\label{eqn:human_action}
\end{equation}
where the second and third set of terms indicate the interaction among the various effects of the human action on the PB parameters. Of course, higher order interactions terms can be considered, but we shall restrict our considerations up to second order and, in fact, to a subset of planetary boundary parameters. It is physically reasonable and mathematically convenient to assume that the $9 \times 9$ matrix, $[g_{ij}]$ is symmetric, $g_{ij} = g_{ji}$, and non-degenerate, $\det[g_{ij}] \neq 0$. In principle, these interactions terms are sub-dominating, however, their importance has to be established empirically. As discussed in Ref.\,\cite{Bertolami:2018b}, they can lead to new equilibria and suggest some mitigation strategies depending on the sign of the matrix entries, $g_{ij}$, and their strength \cite{Bertolami:2018b}.  

Indeed, in order to understand these possibilities, consider only a couple of parameters, say $h_1$ and $h_9$, and assume, in particular, that the ninth parameter corresponds to the Technosphere \cite{Gaffney:2017}, \textit{i.e.}, the set of human technological activities aiming to repair or to mitigate the action on the variables away from the SOS , as suggested in Ref.\,\cite{Bertolami:2018b}. Hence, we can write $H$ as,
\begin{equation}
	H = h_1 + h_9 + 2 g_{19} h_1 h_9 + g_{11} h_1^2 + g_{99} h_9^2.
\end{equation}
It is easy to see that if $h_1 > 0$, $h_9 < 0$ and $g_{19} > 0$, then the effect is to mitigate the destabilizing effect of $h_1$. The net effect of this technological interaction is to ensure that the minimum due to human intervention is, as discussed in Ref.\,\cite{Bertolami:2018a}, closer to the Holocene, minimizing the departure of the ES temperature from the one at the Holocene. We could also argue that $g_{11}$ and $g_{99}$ might be negative too, given that they can have a saturating effect on themselves.


\section{
	The Accounting Framework 
	\label{sec:TheAccountingFramework}
}

There is a natural accountancy criterion built into the physical framework discussed in the previous section, given that it allows for the quantification of the destabilizing effect of human intervention on the ES. This can be exemplified when we consider the depletion of living biomass stock, as discussed in Ref.\,\cite{Bertolami:2018a}. Considering this as the sole component of human intervention,
\begin{equation}
\label{eqn:HLB}
	H = \alpha \Delta m_{\rm LB},
\end{equation}
where $m_{\rm LB}$ is the total living biomass and $\alpha = 3.5 \times 10^7\,{\rm J/kg}$ is the conversion constant of biomass into energy. Estimates indicate that the amount of living biomass has dropped from $750 \times 10^{12}\,{\rm kg}$ in 1800, to $660 \times 10^{12}\,{\rm kg}$ in 1900, and to $550 \times 10^{12}\,{\rm kg}$ in 2000 \cite{Smil:2011,Bar-On:2018}.

This biomass depletion is well correlated with, {\it e.g.} $\rm CO_2$ concentration, hence it suggests a quota, $Q_{BM}$, for a country or region at a fixed time interval, $\Delta t$, by multiplication by a fraction, $f({\rm A,P,GDP,\dots})$, of the global quota that is a function of the territory's area, population, GDP or other relevant measures,
\begin{equation}
	Q_{\rm LB} = \frac{\alpha \Delta m_{\rm LB}}{\Delta t} f({\rm A,P,GDP,\ldots}),
\end{equation}
Of course, the specific planetary boundary breakdown suggests itself a set of quotas:
\begin{equation}
	Q_{h_i} = {h_i \over \Delta t} f({\rm A,P,GDP,\ldots}),
\end{equation}
which requires an accurate knowledge of all $h_i$‘s and their scaling down to a country (region). Furthermore, the usefulness of this approach is valid as far as the contribution of the interaction terms is sub-dominant at the time interval $\Delta t$, otherwise, this set of quotas is meaningless in this time interval.  

The temperature field can have, as discussed in Ref.\,\cite{Bertolami:2018a}, a spatial dependence and thus the free energy should have terms like $|\nabla \psi|^2$ and $\nabla^2 \psi$. These terms could be split in contributions from countries (regions) area and hence should be normalized by their area. This suggests the following quota systematics for a given country (region):
\begin{equation}
	Q_{\rm Spatial} = {|\nabla \psi|^2 \over \Delta t} f(A),
\end{equation}
or an equivalent expression for the Laplacian.

Hence it is clear that our physical framework suggests several strategies for an accountancy procedure. Neglecting interactions, the breakdown of H into its planetary boundary components leads, for a particular choice of the fraction, $f(A,P,GDP,\ldots)$ to the Planetary Boundary Accounting Framework of Ref. \cite{Meyer:2018} up to a multiplicative constant that converts the matter concentration and flows of each of the planetary boundary quantities into the free energy involved in the respective set of thermodynamic processes. For sure, these processes must be understood and decomposed in its most elementary steps so that their very essence is captured and quantified.

In fact, it is fairly easy to show that our procedure encompasses other accounting proposals through a specific choice of the subset of terms of the breakdown of H in its Planetary Boundary (PB) components. For instance, if we consider the Technosphere, as discussed above, it is reasonable to assume that its impact must be conjugated with the population, $P$, and affluence, $A$, or GDP, which implies that a particular term of our breakdown of H contains the $I=PAT$ proposal. Of course, since the $I=PAT$ measure is purely socio-economical, for instance, $T$ is measured in the number of patents, it is somewhat disjoint from the thermodynamical nature of our approach, but there is a clear parallelism of the two measures once a suitable planetary boundary is chosen. 

It is relevant to point out that none of the accounting systems proposed so far includes the interactions terms and some among these might be clearly important. If so, the next issue is to consider the typical time scale for a given planetary boundary parameter to affect the others. For instance, in what concerns the relationship between the $\rm CO_2$ concentration and its effect on the acidity of the oceans a typical time scale of change is about 240 days \cite{Zeebe:2001}. The understanding of the underlying processes associated to this change will be discussed in the following section. If this time scale is the smallest one involving the interaction between the planetary boundary parameters, then the accounting system of Ref. \cite{Meyer:2018} yields a reliable picture of the ES at this time scale. 


\section{Modelling the interaction between Planetary Boundaries}

As discussed in the previous section, for each PB, there is a control variable that measures its deviation from the Holocene conditions. Each of these control variables, $x_i$ can then be associated to an $h_i$ term in the human activities function given by Eq.\,(\ref{eqn:human_action}) by means of a proportionality constant $\alpha_i$. The correspondence is thus given by,
\begin{equation}
	h_i = \alpha_i \Delta x_i.
\end{equation}
The $\alpha_i$ proportionality constants reflect the physical processes of each PB and have units such that the $h_i$ terms have dimensions of energy.

We present here the modelling of the interaction between climate change and the ocean acidification. We will denote these with indexes $i=1,2$, respectively, and the interaction term between them by $g_{12}$. 

In both these cases, the control variables, $x_1,x_2$, have units of concentration, namely, Carbon Dioxide concentration in the atmosphere, for climate change, and $\rm H^{+}$ ion concentration in seawater, for ocean acidification.

These choices allows for writing the PB in the following way:
\begin{align}
	h_1 &= \alpha_1 \Delta x_1 = \alpha_1(x_1 - x_{1,\rm H} + \epsilon), \label{eqn:h1} \\
	h_2 &= \alpha_2 \Delta x_2 = \alpha_2(x_2 - x_{2,\rm H}),
\end{align}	
where $\epsilon$ is the direct anthropogenic carbon emission in a given time period and $x_{i,\rm H}$ correspond to Holocene values.

To calculate the interaction term $g_{12}$ we examine two different scenarios, the non-interacting and the interacting ones. In the first scenario, since there is no interaction there is no change in $h_{2}$ due to carbon dioxide concentration increase, so the global variation is equal to the change in $h_{1}$.
\begin{equation}
	\Delta H = \alpha_{1} (x_{1} - x_{1,\rm H} + \epsilon) + \alpha_{2}(x_{2} - x_{2,\rm H}).
\end{equation}
On the other hand, if we consider $g_{12}\neq0$, it is known that the variation of $H$ over the time scale long enough to achieve a quasi-static equilibrium is

\begin{align}
	\Delta H &= \alpha_1 (x_1 - x_{1,\rm H}) + \alpha_2 (x_2 - x_{2,\rm H}) + \nonumber \\
	& \quad + \alpha_1 \delta x_{1} + \alpha_2 \delta x_{2},
\end{align}

where $\delta x_1$ and $\delta x_2$ are changes in the concentrations of carbon and hydrogen ion respectively, due to the existence of interaction. We shall compute these quantities later.
All factors considered, the proposition is that if we add the interaction term to the equation that resulted from the non-interaction scenario we should get the equation from the interaction scenario, yielding:
\begin{equation}
	\Delta H(g_{12}=0) + 2g_{12}h_{1}h_{2} = \Delta H(g_{12}\neq0),
\end{equation}
which yields an equation for the value of the interaction term,
\begin{equation}
	\epsilon + 2g_{12}h_1 h_2 = \alpha_{1}\delta x_{1} + \alpha_{2}\delta x_{2},
\end{equation}
that we can solve to obtain an expression for the interaction term, $g_{12}$, from empirical data,
\begin{equation}
	g_{12} = \frac
		{\alpha_1 (\delta x_1 - \epsilon) + \alpha_{2} \delta x_2}
		{2 \alpha_1 \alpha_2 (x_1 - x_{1,\rm H} + \epsilon) (x_2 - x_{2,\rm H})}.
	\label{eqn:interactionTerm}
\end{equation}
The previous equation shows that the existence of an interaction between the carbon dioxide in the atmosphere and the ocean acidity leads to perturbations, $\delta x_i$'s, of the control variable. But we still need to develop a method that yields these variations as an output given the annual carbon emission values as an input. It is here that the knowledge of the carbonate system is particularly important.

First, the relation between the concentration of $\rm CO_2$ in the atmosphere and the oceans is regulated by a chemical equilibrium, with a solubility constant $K_0$,
\begin{equation}
	{\rm CO_2\,(g.)}  \leftrightharpoons {\rm CO_2\,(aq.)}.
\end{equation}

Then, in the ocean, there is a well known chain of chemical reactions that describes the carbonate system and may be broken down into two simple reactions,
\begin{align}
	{\rm CO_2 + H_2 O} & \underset{\bar{k}_1}{\stackrel{k_1}{\leftrightharpoons}} {\rm HCO_3^{-} + H^{+} }, \label{eqn:chemCO2} \\
	{\rm HCO_3^{-}} & \underset{\bar{k}_{2}}{\stackrel{k_{2}}{\leftrightharpoons}} {\rm CO_3^{2-} + H^{+}}, \label{eqn:chemCO3-}
\end{align}
where the $k_{i}$ are the forward reaction rate and $\bar{k}_{i}$ the reverse reaction rate coefficients. We now discuss the dynamics involved in the carbonate system and how the increase of carbon dioxide in the atmosphere due to anthropogenic carbon emissions leads to the acidification of the ocean. 

First we need to describe mathematically how the concentrations of all inorganic forms in the ocean vary in time after the system's equilibrium is disturbed, that is, when there is a gradual injection of carbon dioxide into the system. Recalling Eqs. (\ref{eqn:chemCO2}) and (\ref{eqn:chemCO3-}), and introducing the following notation, $v = [{\rm CO_2}]$, $w = [{\rm H^+}]$, $y = [{\rm HCO_3^-}]$, $z = [{\rm CO_3^{2-}}]$, then:
\begin{align}
	\frac{d v}{dt} &= - k_1 v + \bar{k}_1 y w, \label{eqn:kineticCO2} \\
	\frac{d y}{dt} &= k_1 v - \bar{k}_1 y w - k_2 y + \bar{k}_2 z w, \\
	\frac{d z}{dt} &= k_2 y - \bar{k}_2 z w, \\
	\frac{d w}{dt} &= k_1 v - \bar{k}_1 y w + k_2 y - \bar{k}_2 z w.
\end{align}
Note that, above, we defined $x_1$ as the ${\rm CO_2}$ concentration in the atmosphere and assigned it as the control variable for climate change. Now, we are using $v$ as the ${\rm CO_2}$ concentration in the oceans. Since these two concentrations follow an equilibrium dictated by a solubility constant, they can be interchanged as the control variable for climate change. 

We point out that Eq. (\ref{eqn:kineticCO2}) is the time evolution of carbon dioxide's concentration simply due to the chemical dynamics, but it is of our interest to study how the entire system reacts if there is a continuous input of carbon dioxide, even though by small quantities at a time. From Ref.\,\cite{Carbon:a} we obtain that over the last year (2018) the average concentration of carbon dioxide in the atmosphere has been increasing almost linearly\footnote{This linear approximation might be a crude one, but it is still interesting to see the outcome of this simple scenario.}, with a rate of $0.250$ ppm per month (from 407 ppm in January 2018 to 410 ppm in January 2019).

In Ref.\,\cite{Zeebe:2001}, it is estimated that the time it takes for the exchange between gaseous and aqueous carbon dioxide to reach an equilibrium is about a year (240 days). With that in mind, we consider that at the start of an year (e.g January 2018) carbon dioxide exchange has reached an equilibrium and an atmospheric concentration of currently $407$ ppm, and consider a steady input of carbon such that at the start of the following year (in this scenario January 2019) the increase in concentration was of $3$ ppm granting a total of $410$ ppm, meaning that there was an increase of about $0.737\%$ of $\rm CO_2$ in the atmosphere. Over the year the oceans must have reached a new equilibrium and since the percentual increase was small, the new equilibrium should not be too far away from the old one. Taking this into account, it seems reasonable to consider a Linear Stability approach to compute how the concentrations of the inorganic forms respond to a steady, but small increase of $CO_{2}$ in the system. Defining a vector $\vec{r}$ as: 
\begin{equation} 
	\vec{r}=
	\begin{pmatrix}
		v \\
		y \\
		z \\
		w
	\end{pmatrix},
\end{equation}
and a small perturbation around a fixed $\vec{r}$, $\vec{r}\rightarrow\vec{r}+\delta\vec{r}$. From the linear stability approach $\dot{\delta r_{i}}=\frac{\partial\dot{r_i}}{\partial r_j}|_{\vec{r}_0}\,\delta r_{j} + \mathcal{O}(\delta r^{2})$, where the dot stands for the usual time derivative and $\vec{r}_0$ is the fixed point for which $\delta\vec{r}=0$. We may represent the derivative terms as components of a matrix $M$ such that $M_{ij}=\frac{\partial\dot{r_{i}}}{\partial r_{j}}|_{\vec{r}_0}$. The equilibrium condition gives the following concentrations:
\begin{equation}
	\vec{r}_0 =
		\begin{pmatrix}
			v_0 \\
			y_0 \\
			z_0 \\
			w_0
		\end{pmatrix}
	=
		\begin{pmatrix}
			v_0 \\
			\frac{k_{1}}{\bar{k}_{1}}\frac{v_0}{w_0} \\
			\frac{k_{1}k_{2}}{\bar{k}_{1}\bar{k}_{2}}\frac{v_0}{{w_0}^{2}} \\
			w_0
		\end{pmatrix}
\end{equation}
Given these values the matrix takes the form:
{\footnotesize
\begin{equation} M=
	\begin{pmatrix}
		-k_1 	& \bar{k}_1 w_0 	& 0 	& k_1 \frac{v_0}{w_0} \\
		k_1 	& -\bar{k}_1 w_0 - k_2 	& \bar{k}_2 w_0 	& -k_{1}\frac{v_0}{w_0} + \frac{k_1 k_2}{\bar{k}_2} \frac{v_0}{{w_0}^2} \\
		0 		& k_2 	& -\bar{k}_2 w_0 & -\frac{k_1 k_2}{\bar{k}_2} \frac{v_0}{{w_0}^2} \\
		k_1 	& -\bar{k}_1 w_0 + k_2 		& -\bar{k}_2 w_0 	& -k_1 \frac{v_0}{w_0} - \frac{k_1 k_2}{\bar{k}_2}\frac{v_0}{{w_0}^2}
	\end{pmatrix}
\end{equation}}
That is written in terms of $v_0$ and $w_0$ which can be calculated by knowing the concentration of carbon dioxide ($v_0$) and seawater $\rm pH$ ($w_0$) at a given time. The dynamical equations around the equilibrium point are then given by:
\begin{equation}
	\dot{\delta\vec{r}}=M\delta\vec{r}+m\hat{e}_{x},
	\label{eqn:dynamical}
\end{equation}
where the term $m\hat{e}_{x}$ is the external contribution from the steady increase in the $\rm CO_2$ concentration.


\section{Results}

Even though we have already considered an approximation for the dynamics, it is still hard to calculate the expressions for the equilibrium concentrations, therefore in this section we shall consider some numerical values and interpret the results in terms of these particular values.

First let us take into account the amount of carbon dioxide that was emitted during 2017 and 2018. It is estimated in Ref.\,\cite{Carbon:b}, that in 2017 approximately $36.1 \times 10^{12}\,{\rm kg}$ were emitted, and in 2018 approximately $37.1 \times 10^{12}\,{\rm kg}$. It is possible to convert these quantities into concentrations in the atmosphere. The conversion rate is $1\,{\rm ppm(CO_2)} = 2.12 \times 10^{12}\,{\rm kg}$. We should note that in these two consecutive years there was an increase of approximately $2.5\%$ in carbon emission, meaning that, if this trend continues, by the end of 2019 the total amount of carbon emitted will be close to $38.0 \times 10^{12}\,{\rm kg}$. It is helpful to summarize some relevant quantities that can be derived from the previous values.

\begin{table}[h]
	\caption{Carbon dioxide annual emission and related values. Starred values ($*$) are predictions based on the extrapolation of carbon emissions.}
	\label{tbl:CO2Emission}
	\centering
	\begin{tabular}{>{\raggedright}p{4cm}|c c c}
		Year 															& 2017 	& 2018 	& 2019 \\
		\hline
		\hline
		$\rm CO_2$ emissions ($10^{12}\,{\rm kg}$) 						& 36.2 	& 37.1 	& 38.0{*} \\
		\hline 
		Atmospheric $\rm CO_2$ equilibrium concentration, $x_1$  (ppm) 	& 406 	& 407 	& 410 \\
		\hline 
		Oceans $\rm CO_2$ equilibrium concentration Jan.\,1st, $v_0$ ($10^{-5}\,{\rm mol/dm^3})$ 
																		& 1.218 & 1.221 & 1.230 \\
		\hline 
		Atmospheric $\rm CO_2$ concentration increase, $\epsilon$ (ppm)	& 17.1 & 17.5 & 17.9{*} \\
		\hline 
		Oceans $\rm CO_2$ concentration increase, $m$ $(10^{-14}\,{\rm mol/(dm^3s})$ 
																		& 1.62 & 1.66 & 1.70{*}
	\end{tabular}
\end{table}

The molar concentrations in Table\,\ref{tbl:CO2Emission} can be calculated using Dalton's Law for partial pressure and the value for the solubility constant, $K_{0}$, was estimated using the information provided in Ref.\,\cite{Zeebe:2001} assuming a temperature of $298 K$, which gives a value for the solubility constant $K_{0}\approx0.03\,{\rm mol/(dm^3atm)}$.

Let us now verify whether our model is consistent and if we can forecast some values for the beginning of 2020. The average seawater $\rm pH$ in 2017 may have been very close to $8.07$, \cite{Zeebe:2001}, which indicates an hydrogen ion concentration of $8.44 \times 10^{-9}\,{\rm mol/dm^3}$. Recalling the matrix $M$ and the concentrations' vector, $\vec{r}$, described in the previous section, it is clear that in order to calculate the variations of all components' concentrations we need the numerical values for the equilibrium concentrations of $\rm [CO_{2}]$, $v_0$, and $\rm [H^{+}]$, $w_0$, as well as the forward and reverse reaction rate constants. The equilibrium concentrations $v_0$ are given in Table\,\ref{tbl:CO2Emission} and $w_0$ is the value calculated from the assumed $\rm pH$, namely, $8.07$. The change per unit of time, $m$, is also shown in Table\,\ref{tbl:CO2Emission}. The reaction rates are provided in Ref.\,\cite{Schulz:2006} and are the following:

\begin{align}
	k_1 &= 3.71 \times 10^{-2}\,{\rm s^{-1}}, \\
	\bar{k}_1 &= 2.67 \times 10^4\,{\rm dm^3/(mol\,s)}, \\
	k_2 &= 59.44\,{\rm s^{-1}}, \\
	\bar{k}_2 &= 5.0 \times 10^{10}\,{\rm dm^3/(mol\,s)} .
\end{align}

According to the definition of the PB's variables, we may identify, up to a constant, $v$ with the atmospheric carbon concentration, $x_1$, and $w$ with the hydrogen ion concentration in the ocean $x_2$. Now we substitute the values into Eq.\,(\ref{eqn:dynamical}) and consider a time scale of one year, substituting the time, $t$, by $3.16\times10^{7}s$. Starting from 2017 with $v_0 = 1.22\times10^{-5}$, $w_0 = 8.44 \times 10^{-9}$ and $m = 1.62 \times 10^{-14}$, the values for the variations in the concentrations $v$ and $w$ are:
{\setlength\arraycolsep{2pt}
\begin{eqnarray}
	\delta v(2017 \rightarrow 2018) &=& 3.24\times10^{-8}, \\
	\delta w(2017 \rightarrow 2018) &=& 1.84\times10^{-11}.
\end{eqnarray}}

According to the proposed model the concentration $x$ at the beginning of an year should be the concentration at the start of the previous year $v_0$ (assuming that equilibrium was achieved at the time) plus the small variation $\delta v$. So, if in 2017 we start with $v_0 = 1.218 \times 10^{-5}$, then in 2018 we should have:
\begin{equation}
	v_0(2018) = v_0(2017) + \delta v = 1.221 \times 10^{-5} .
\end{equation}

From Table\,\ref{tbl:CO2Emission} we see that in 2018 $v_0$ was $1.22\times10^{-5}$ which is fairly close to the value given in Eq. (35). Also, we get a new value for $w_0$ by adding $\delta w$:
\begin{equation}
	w_0(2018) = w_0(2017) + \delta w = 8.46 \times 10^{-9},
\end{equation}
which gives a new value for the acidity level, $pH=8.073$. A decrease of $0.0124\%$. 

Now let us do the same for the time period between 2018 and 2019, $v_0=1.221 \times 10^{-5}$, $w_0 = 8.461 \times 10^{-9}$ and $m=1.66 \times 10^{-14}$:

\begin{align}
	\delta v(2018 &\rightarrow 2019) = 3.33 \times 10^{-8}, \\
	\delta w(2018 &\rightarrow 2019) = 1.89 \times 10^{-11}.
\end{align}

From these values we can perform the same kind of computation as before
to obtain:
\begin{equation}
	v_0(2019) = v_0(2018) + \delta v = 1.224\times10^{-5},
\end{equation}
which once again is very close to the value given in Table\,\ref{tbl:CO2Emission}, $v_0=1.230 \times 10^{-5}$ up to a $0.49\%$ increase. The new value of $w_0$ is:
\begin{equation}
	w_0(2019) = w_0(2018) + \delta w = 8.48 \times 10^{-9},
\end{equation}
which gives a new value for the acidity level, ${\rm pH} = 8.072$, a decrease
of $0.0124\%$.

So if we consider the values shown in Table\,\ref{tbl:CO2Emission} for the carbon emissions during 2019, and recalling that those values are assuming the same increasing trend in the amount of emissions, it is possible to forecast values for the following year 2020. Then, for the time period between 2019 and 2020: $v_0 = 1.23 \times 10^{-5}$, $w_0 = 8.48 \times 10^{-9}$, $m=1.70 \times 10^{-14}$:
\begin{align}
	\delta v(2019 \rightarrow 2020) &= 3.42 \times 10^{-8}, \\
	\delta w(2019 \rightarrow 2020) &= 1.93 \times 10^{-11}.
\end{align}
Leading to the following equilibrium values for 2020:
\begin{equation}
	v_0(2020) = v_0(2019) + \delta v = 1.233 \times 10^{-5},
\end{equation}
assuming an $0.49\%$ error  in the previous calculation. This value of $v_0$ means that by the end of the year (2019) and the beginning of 2020 the carbon dioxide concentration in the atmosphere would be approximately $(411 \pm 2)\,{\rm ppm}$. For the hydrogen ion we get:
\begin{equation}
	w_0(2020) = w_0(2019) + \delta w = 8.50 \times 10^{-9} ,
\end{equation}
which gives a new value for the acidity level, ${\rm pH} = 8.07$, corresponding to a decrease of $0.0124\%$. The same calculations were performed for each year of the last decade and the results are summarized in Table\,\ref{tbl:AnualCO2}.

\begin{table*}[htp]
	\caption{Carbon dioxide annual emission and related values}
	\label{tbl:AnualCO2}
	{\footnotesize
	\begin{tabularx}{\textwidth}{>{\raggedright}X|c|c|c|c|c|c|c|c|c|c|c}
		{Year} & {2010} & {2011} & {2012} & {2013} & {2014} & {2015} & {2016} & {2017} & {2018} & {2019} & {2020} \\
		\hline
		\hline
		$\rm CO_2$ emissions ($10^{12}\,{\rm kg}$)
		& {33.1} & {34.4} & {35.0} & {35.3} & {35.6} & {35.5} & {35.7} & {36.2} & {37.1} & {38.0{*}} & {39.0{*}} \\
		\hline 
		Atmospheric $\rm CO_2$ concentration increase, $\epsilon$ (ppm)
		& {15.6} & {16.2} & {16.5} & {16.7} & {16.8} & {16.8} & {16.8} & {17.1} & {17.5} & {17.9{*}} & {18.38{*}} \\
		\hline 
		Atmospheric $\rm CO_2$ equilibrium concentration, $x_1$ (ppm)
		& {388} & {391} & {393} & {395} & {397} & {399} & {402} & {406} & {407} & {410} & {413*} \\
		\hline 
		Oceans $\rm CO_2$ equilibrium concentration, $v_0$ ($10^{-5}\,{\rm mol/dm^3})$
		& {1.164} & {1.173} & {1.179} & {1.185} & {1.191} & {1.197} & {1.206} & {1.218} & {1.221} & {1.230} & {1.233{*}} \\
		\hline 
		Oceans $\rm CO_2$ concentration increase, $m$ $(10^{-14}\,{\rm mol/(dm^3s})$
		& {1.48} & {1.54} & {1.57} & {1.58} & {1.59} & {1.59} & {1.60} & {1.62} & {1.66} & {1.70{*}} & {1.75{*}} \\
		\hline 
		Oceans $\rm H^{+}$ equilibrium concentration, $w_0$ $(10^{-9}\,{\rm mol/dm^3})$
		& {8.32} & {8.34} & {8.35} & {8.37} & {8.39} & {8.41} & {8.43} & {8.44} & {8.46} & {8.48} & {8.50{*}} \\
		\hline 
		{$\delta v~(10^{-8}{\rm mol\,dm^{-3}})$} 
		& {2.89} & {3.02} & {3.08} & {3.12} & {3.15} & {3.15} & {3.18} & {3.24} & {3.33} & {3.42{*}} & {3.53{*}} \\
		\hline 
		{$\delta w~(10^{-11}{\rm mol\,dm^{-3}})$} 
		& {1.69} & {1.75} & {1.79} & {1.80} & {1.82} & {1.81} & {1.82} & {1.84} & {1.89} & {1.93{*}} & {1.99{*}} \\
		\hline 
		{New oceans $\rm CO_2$ concentration, $v_0 + \delta v$ $(10^{-5}\,{\rm mol/dm^3})$} 
		& {1.167} & {1.176} & {1.182} & {1.188} & {1.194} & {1.200} & {1.210} & {1.221} & {1.224} & {1.233{*}} & {1.240{*}} \\
		\hline 
		{New oceans $\rm H^{+}$ concentration, $w_0 + \delta w$ $(10^{-9}\,{\rm mol/dm^3})$} 
		& {8.34} & {8.35} & {8.37} & {8.39} & {8.41} & {8.43} & {8.44} & {8.46} & {8.48} & {8.50{*}} & {8.52{*}} \\
		\hline 
		{New atmosphere $\rm CO_2$ concentration, $x_1 + \delta x_1$ (ppm)} 
		& {389} & {392} & {394} & {396} & {398} & {400} & {403} & {407} & {408} & {411{*}} & {415{*}} \\
		\hline 
		{New oceans pH} 
		& {8.079} & {8.078} & {8.077} & {8.076} & {8.075} & {8.074} & {8.074} & {8.073} & {8.072} & {8.071{*}} & {8.071{*}} \\
		\hline 
		{Atmosphere $\rm CO_2$ concentration error} 
		& {0.52\%} & {0.25\%} & {0.25\%} & {0.25\%} & {0.25\%} & {0.50\%} & {0.66\%} & {\textasciitilde 0} & {0.49\%} & - & - \\
	\end{tabularx}
	}
\end{table*}

To determine the interaction term, $g_{12}$, we return to Eq.\,(\ref{eqn:interactionTerm}), for which we now have all the values, except for $\alpha_1$ and $\alpha_2$. Since $\alpha_1$, $\alpha_2$ and $g_{12}$ should reflect the complex dynamics of the ES processes, they would, at least theoretically, be constant. Since, we are only looking at the interaction between two processes and are describing them perturbatively, we should expect to be able to find a reasonable approximation for the relationships between these three parameters from the data in Table\,\ref{tbl:AnualCO2}.

\begin{figure}
	\centering
    \includegraphics[width=\columnwidth]{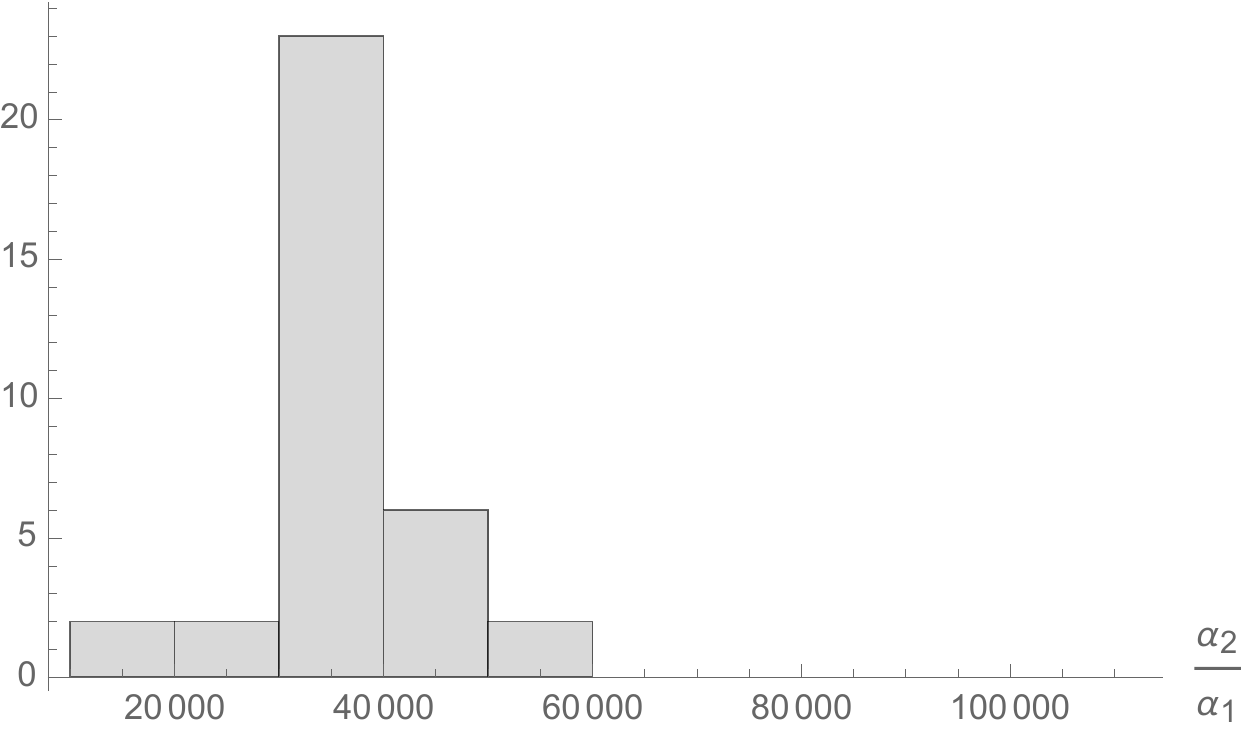}
	\caption{Histogram of the sample of values for the $\alpha_2/\alpha_1$ ratio from the data resulting from the years 2010 to 2018. The mean value is $\langle \alpha_2/\alpha_1 \rangle = 57056$.}
	\label{fig:distributionAlphaRelation}
\end{figure}

Recalling Eq.\,(\ref{eqn:interactionTerm} and using the data from Table\,\ref{tbl:AnualCO2}, we can write multiple equations for $g_{12}$ by assuming that $g_{12}$ is constant and taking any pair of years and equating their correspondent equation, \textit{e.g.} $g_{12}(2010)=g_{12}(2011)$. Doing this equality for all available combinations yields a sampling of experimental values for the proportionality constant $\alpha_2/\alpha_1$, the results of which are depicted in Fig.\,\ref{fig:distributionAlphaRelation}. 

With this relationship, we can then write $g_{12}$ as a function of only one of them. At first glance, it seems more obvious to write it as a function of $\alpha_1$,
\begin{equation}
	g_{12} = \frac{A}{\alpha_1},
	\label{eqn:interactionTermA}
\end{equation}
because $\alpha_1$ is the constant related to the greenhouse mechanism in climate change, as defined in Eq.\,(\ref{eqn:h1}).

\begin{figure}
	\centering
    \includegraphics[width=\columnwidth]{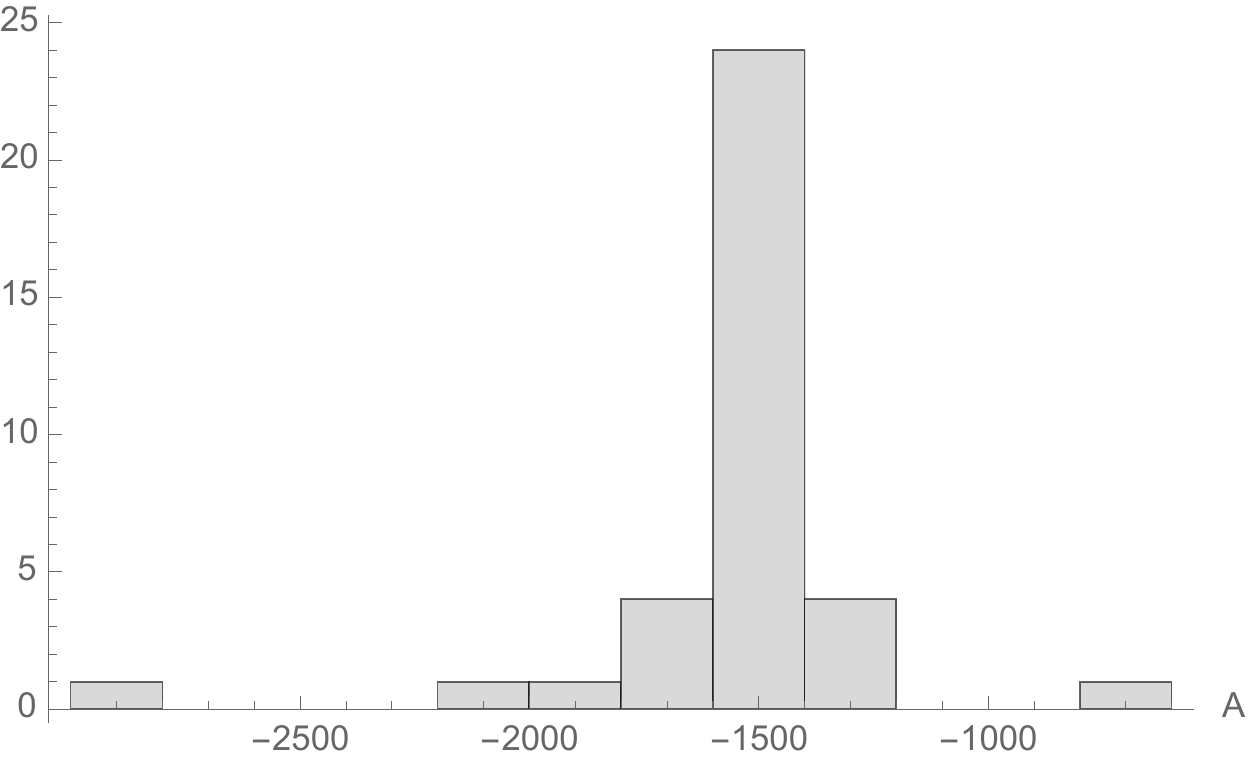}
	\caption{Histogram for the $A$ constant that determines the interaction term $g_{12}$ in terms of $\alpha_1$, as per Eq.\,(\ref{eqn:interactionTermA})}
	\label{fig:distributionA}
\end{figure}

To each value of $\alpha_2/\alpha_1$ corresponds a values of $A$, which leads to the distribution depicted in Figure\,\ref{fig:distributionA}, with a mean value of $-1548.7\,{\rm mol^{-1}dm^3}$ and a standard deviation of $307.1\,{\rm mol^{-1}dm^3}$. The fact that the individual results seem to converge to a single value is reassuring that the assumptions made so far are reasonable.

All that remains now is to determine $\alpha_1$. A full determination would require a knowledge of the $a(\eta)$ and $b(\eta)$ functions that control natural drivers of the ES in Eq.\,(\ref{eqn:FreeEnergy}). Still, we can use the biomass accounting described in Section\,\ref{sec:TheAccountingFramework} as a rough estimate for the purposes of getting orders of magnitude for the different terms. Thus, we take the average annual rate of  biomass depletion at $10^{13}\,{\rm kg/year}$, of which around $10^{12}\,{\rm kg/year}$ is living biomass, and its energy content $3.5 \times 10^7\,{\rm J/kg}$. The conversion of living biomass into carbon dioxide emissions is not straightforward and depends on its exact composition, but if we assume that carbon atoms make up for the majority of the mass of organic matter, by comparing the molar mass of carbon and carbon dioxide, we can estimate that there will be $3.6\,{\rm kg}$ if $\rm CO_2$ emitted for each $\rm kg$ of biomass consumed, owing to the incorporation of oxygen in combustion, that matches the values in Table\,\ref{tbl:CO2Emission}.

We can use this to obtain a figure of merit for $\alpha_1 \sim 4 \times 10^{26}\,{\rm J/(mol\,dm^3)}$ that relates the increase in carbon dioxide concentration in the atmosphere with energy degradation. We can then use the mean value for the relationship between $\alpha_1$ and $\alpha_2$ obtained as described above. This means that $\alpha_2 \sim 6 \times 10^4 \alpha_1 = 2.4 \times 10^{31}\,{\rm J/(mol\,dm^3)}$ and $g_{12} \sim 3.4 \times 10^{-24}\,{\rm J^{-1}}$. These orders of magnitude, by themselves, have little meaning, since they are comparing different things. However, we can compare the relative importance of each $h_i$ contribution and the interaction term, $g_{12}$. Thus,
\begin{align}
	h_1 &= \alpha_1 \Delta x_1 \sim 4 \times 10^{26} \times 5 \times 10^{-6} = \nonumber \\ 
	&= 2 \times 10^{21}\,{\rm J} \\
	h_2 &= \alpha_2 \Delta x_2 \sim 2 \times 10^{31} \times 2 \times 10^{-9} = \nonumber \\ 
	&= 4 \times 10^{22}\,{\rm J} \\
	g_{12} h_1 h_1 &\sim - 3.4 \times 10^{-24} \times 2 \times 10^{21} \times 4 \times 10^{22} = \nonumber \\
	&= 2.7 \times 10^{20}\,{\rm J}
\end{align}

The main takeout from these results is that the interaction term is at least one order of magnitude below the other two terms, which mean that it is relevant but should not affect the overall consistency of an accounting system within the time scale of a year or so.


\section{Conclusions}

In this work we have shown that the physical description of the ES in terms of the theory of phase transitions of Landau-Ginzburg discussed in Refs. \cite{Bertolami:2018a, Bertolami:2018b} provides a natural accounting framework for measuring the impact of the human drivers once these are broken in terms of planetary boundary components.

The arising framework suggests a few accounting strategies, which can be gauged in terms of the population, the area, or the GDP of a given country or region. We have shown how a quota system can be built from the particular example of the depletion of living biomass. A similar systematics would lead to a quota system based on the planetary boundaries and it is argued  that it closely resembles the quota system of Ref.\,\cite{Meyer:2018}. 

Furthermore, we have discussed the role of the interaction terms between the planetary boundary parameters and how they render a quota system valid in a time scale typically closer than the slowest interacting process. We worked out a specific example involving the interaction of the $\rm CO_2$ concentration and the ocean acidity establishing the procedure to obtain a description of human action that can have interaction terms. This theoretical exercise is an important first step in achieving a useful modelling procedure for the ES components that can be then inserted into a dynamical description of the ES that can establish the conditions under which it can remain within the Safe Operating Space.


\section*{Acknowledgements}

\noindent
The authors would like to thank Kate Meyer, Will Steffen and Alessandro Galli for the insightful discussions. The work of FF is supported by the Fundação para a Ciência e Tecnologia through grant SFRH/BPD/118649/2016.


\bibliographystyle{unsrt}

\bibliography{accounting_system}

\end{document}